\documentclass[12pt]{article}
\usepackage{graphicx}
\usepackage{lscape}
\usepackage{citesort}
\usepackage{amssymb}
\usepackage{appendix}
\usepackage{multirow}

\begin{document}
\def\qq{\langle \bar q q \rangle}
\def\uu{\langle \bar u u \rangle}
\def\dd{\langle \bar d d \rangle}
\def\sp{\langle \bar s s \rangle}
\def\GG{\langle g_s^2 G^2 \rangle}
\def\Tr{\mbox{Tr}}
\def\figt#1#2#3{
        \begin{figure}
        $\left. \right.$
        \vspace*{-2cm}
        \begin{center}
        \includegraphics[width=10cm]{#1}
        \end{center}
        \vspace*{-0.2cm}
        \caption{#3}
        \label{#2}
        \end{figure}
	}
	
\def\figb#1#2#3{
        \begin{figure}
        $\left. \right.$
        \vspace*{-1cm}
        \begin{center}
        \includegraphics[width=10cm]{#1}
        \end{center}
        \vspace*{-0.2cm}
        \caption{#3}
        \label{#2}
        \end{figure}
                }

\def\ds{\displaystyle}
\def\beq{\begin{equation}}
\def\eeq{\end{equation}}
\def\bea{\begin{eqnarray}}
\def\eea{\end{eqnarray}}
\def\beeq{\begin{eqnarray}}
\def\eeeq{\end{eqnarray}}
\def\ve{\vert}
\def\vel{\left|}
\def\ver{\right|}
\def\nnb{\nonumber}
\def\ga{\left(}
\def\dr{\right)}
\def\aga{\left\{}
\def\adr{\right\}}
\def\lla{\left<}
\def\rra{\right>}
\def\rar{\rightarrow}
\def\lrar{\leftrightarrow}  
\def\nnb{\nonumber}
\def\la{\langle}
\def\ra{\rangle}
\def\ba{\begin{array}}
\def\ea{\end{array}}
\def\tr{\mbox{Tr}}
\def\ssp{{\Sigma^{*+}}}
\def\sso{{\Sigma^{*0}}}
\def\ssm{{\Sigma^{*-}}}
\def\xis0{{\Xi^{*0}}}
\def\xism{{\Xi^{*-}}}
\def\qs{\la \bar s s \ra}
\def\qu{\la \bar u u \ra}
\def\qd{\la \bar d d \ra}
\def\qq{\la \bar q q \ra}
\def\gGgG{\la g^2 G^2 \ra}
\def\q{\gamma_5 \not\!q}
\def\x{\gamma_5 \not\!x}
\def\g5{\gamma_5}
\def\sb{S_Q^{cf}}
\def\sd{S_d^{be}}
\def\su{S_u^{ad}}
\def\sbp{{S}_Q^{'cf}}
\def\sdp{{S}_d^{'be}}
\def\sup{{S}_u^{'ad}}
\def\ssp{{S}_s^{'??}}

\def\sig{\sigma_{\mu \nu} \gamma_5 p^\mu q^\nu}
\def\fo{f_0(\frac{s_0}{M^2})}
\def\ffi{f_1(\frac{s_0}{M^2})}
\def\fii{f_2(\frac{s_0}{M^2})}
\def\O{{\cal O}}
\def\sl{{\Sigma^0 \Lambda}}
\def\es{\!\!\! &=& \!\!\!}
\def\ap{\!\!\! &\approx& \!\!\!}
\def\ar{&+& \!\!\!}
\def\ek{&-& \!\!\!}
\def\kek{\!\!\!&-& \!\!\!}
\def\cp{&\times& \!\!\!}
\def\se{\!\!\! &\simeq& \!\!\!}
\def\eqv{&\equiv& \!\!\!}
\def\kpm{&\pm& \!\!\!}
\def\kmp{&\mp& \!\!\!}
\def\mcdot{\!\cdot\!}
\def\erar{&\rightarrow&}

% .........................................................

\def\simlt{\stackrel{<}{{}_\sim}}
\def\simgt{\stackrel{>}{{}_\sim}}

% .........................................................

\renewcommand{\textfraction}{0.2}    %float (figures) parameters
\renewcommand{\topfraction}{0.8}   

\renewcommand{\bottomfraction}{0.4}   
\renewcommand{\floatpagefraction}{0.8}
\newcommand\mysection{\setcounter{equation}{0}\section}

\def\baeq{\begin{appeq}}     \def\eaeq{\end{appeq}}  
\def\baeeq{\begin{appeeq}}   \def\eaeeq{\end{appeeq}}
\newenvironment{appeq}{\beq}{\eeq}   
\newenvironment{appeeq}{\beeq}{\eeeq}
\def\bAPP#1#2{
 \markright{APPENDIX #1}
 \addcontentsline{toc}{section}{Appendix #1: #2}
 \medskip
 \medskip
 \begin{center}      {\bf\LARGE Appendix #1 :}{\quad\Large\bf #2}
% \begin{center}      {\bf\LARGE Appendix  :}{\quad\Large\bf #2}
\end{center}
 \renewcommand{\thesection}{#1.\arabic{section}}
\setcounter{equation}{0}
        \renewcommand{\thehran}{#1.\arabic{hran}}
\renewenvironment{appeq}
  {  \renewcommand{\theequation}{#1.\arabic{equation}}
     \beq
  }{\eeq}
\renewenvironment{appeeq}
  {  \renewcommand{\theequation}{#1.\arabic{equation}}
     \beeq
  }{\eeeq}
\nopagebreak \noindent}

\def\eAPP{\renewcommand{\thehran}{\thesection.\arabic{hran}}}

\renewcommand{\theequation}{\arabic{equation}}
\newcounter{hran}
\renewcommand{\thehran}{\thesection.\arabic{hran}}

\def\bmini{\setcounter{hran}{\value{equation}}
\refstepcounter{hran}\setcounter{equation}{0}
\renewcommand{\theequation}{\thehran\alph{equation}}\begin{eqnarray}}
\def\bminiG#1{\setcounter{hran}{\value{equation}}
\refstepcounter{hran}\setcounter{equation}{-1}
\renewcommand{\theequation}{\thehran\alph{equation}}
\refstepcounter{equation}\label{#1}\begin{eqnarray}}

%       the stuff below defines \eqalign and \eqalignno in such a
%       way that they will run on Latex

\newskip\humongous \humongous=0pt plus 1000pt minus 1000pt
\def\caja{\mathsurround=0pt}
%\def\eqalign#1{\,\vcenter{\openup1\jot
%\caja   %\ialign{\strut \hfil$\displaystyle{##}$&$
%\displaystyle{{}##}$\hfil\crcr#1\crcr}
%}\,}

% ...........................................................

\title{
         {\Large
                 {\bf
Strong coupling constants of heavy spin--3/2 baryons with light pseudoscalar mesons 
                 }
         }
      }

\author{\vspace{1cm}\\
{\small T. M. Aliev$^1$ \thanks {e-mail:
taliev@metu.edu.tr}~\footnote{permanent address:Institute of
Physics,Baku,Azerbaijan}\,\,, K. Azizi$^2$ \thanks {e-mail:
kazizi@dogus.edu.tr}\,\,, M. Savc{\i}$^1$ \thanks
{e-mail: savci@metu.edu.tr}} \\
{\small $^1$ Physics Department, Middle East Technical University,
06531 Ankara, Turkey }\\
{\small$^2$ Department of Physics, Do\u gu\c s University, Ac{\i}badem-Kad{\i}k\"oy,  34722 Istanbul, Turkey}}

\date{}

\begin{titlepage}
\maketitle
\thispagestyle{empty}

\begin{abstract}
The strong coupling constants among  members of the heavy spin--3/2 baryons containing
single heavy quark with light pseudoscalar mesons are calculated
in the framework of the light cone QCD sum rules.
Using symmetry arguments, some structure independent  relations among different 
correlation functions are obtained. It is shown that all possible transitions can be 
described in terms of one universal invariant function whose explicit expression
is Lorenz structure dependent.  
\end{abstract}

%\vspace{1cm}
~~~PACS number(s): 11.55.Hx, 13.75.Gx, 13.75.Jz
\end{titlepage}

\section{Introduction}
The heavy baryons have been at the focus of much attention both theoretically and
experimentally during the last decade. The heavy quarks inside these baryons provide
windows helping us see somewhat further under the skin of the nonperturbative QCD as
compared to the light baryons. Since a part of polarization of heavy quark is
transferred to the baryon, so investigation of polarization effects of these
baryons can give information about the heavy quark spin.
Moreover, these baryons provide a possibility to study the predictions of
heavy  quark effective theory (HQET). Besides the spectroscopy of heavy baryons,
which have been discussed widely in the literature, their  electromagnetic,
weak and strong decays are very promising tools to get knowledge on their
internal structure. In this decade, essential experimental results have been 
obtained in the spectroscopy of heavy baryons. The ${1\over 2}^+$ [ ${1\over 2}^-$] antitriplet states $\Lambda_c^{+},~\Xi_c^{+},~\Xi_c^{0}$ [ $\Lambda_c^{+} (2593)$,
$\Xi_c^{+}(2790),~\Xi_c^{0}(2790)$] as well as the ${1\over 2}^+$ [  ${3\over 2}^+$]  sextet states $\Omega_c,\Sigma_c,\Xi'_c$ [$\Omega_c^\ast,\Sigma_c^\ast,\Xi_c^\ast$] have been observed 
\cite{Rstp01}. Among the S--wave bottom baryons, only the
$\Lambda_b,~\Sigma_b,~\Sigma_b^\ast,~\Xi_b$ and $\Omega_b$ have been discovered.
It is expected that the LHC will open new horizons in the discovery of the excited
bottom baryon sates \cite{Rstp02} and provide possibility to study the electromagnetic 
properties of heavy baryons as well as their weak and strong transitions.
The experimental progress in this area  stimulates intensive theoretical 
studies (for a review see for instance \cite{Rstp03,Rstp04} and references 
therein). Theoretical calculations of parameters characterizing the decay of the 
heavy baryons  will help us  better understand the experimental results.

The strong coupling constants are the main ingredients for strong decays of heavy 
baryons. These couplings occur in a low energy scale  far from the asymptotic 
freedom region, where the strong coupling constant between quarks and gluons is large
and perturbation theory is invalid. Therefore, to calculate such coupling constants 
a nonperturbative approach is needed. One of the most reliable and attractive 
nonperturbative methods is QCD sum rules \cite{Rstp05}. This method is based on QCD 
Lagrangian and does not contain any  model dependent parameter. The light cone  QCD sum 
rules (LCSR) method \cite{Rstp06} is an extended version of the traditional QCD 
sum rules in which the operator product expansion (OPE) is carried over twists rather
than the dimensions of the operators  as in
the case of traditional sum rules. In the present work, we calculate the strong 
coupling constants among sextet of the heavy spin--3/2 baryons containing single 
heavy quark with the light pseudoscalar mesons in the framework of the LCSR.
Using symmetry arguments, we show that all allowed strong transitions among
members of these baryons in the presence of light pseudoscalar mesons can be
expressed in terms of only one universal invariant function. Note that the
coupling constants of heavy spin--1/2 baryons with pseudoscalar and vector mesons
have been recently calculated in \cite{azizi1,azizi2}. The heavy spin 3/2--heavy
spin 1/2 baryon--pseudoscalar meson and  heavy spin 3/2--heavy spin 1/2 baryon--vector
meson coupling constants have also been calculated in the same framework in
\cite{azizi3,azizi4}. It should be mentioned here that the couplings of the
heavy baryons with mesons is first calculated in \cite{R11} 
within the framework of HQET. 

Rest of the  article is organized as follows. In section 2, we derive some structure
independent relations among the corresponding correlation functions, and demonstrate
how the considered coupling constants  can be calculated in terms of only one
universal function. In this section, we also derive the LCSR for the heavy spin--3/2
baryon--light pseudoscalar meson coupling constants using the distribution amplitudes
(DA's) of the pseudoscalar mesons. Section 3 is devoted to the numerical analysis of
the related coupling constants and discussion.

\section{Light cone QCD sum rules for the coupling constants of
pseudoscalar mesons with heavy spin--3/2 baryons}

In this section, the LCSR for the coupling constants among
heavy spin--3/2 baryons and light pseudoscalar mesons are derived. 
We start our discussion by considering the following correlation function:
\bea
\label{estp01}
\Pi_{\mu\nu}= i \int d^4x e^{ipx} \lla {\cal P}(q) \vel {\cal T} \left\{
\eta_{\mu} (x) \bar{\eta}_{\nu} (0) \right\} \ver 0 \rra~,
\eea
where ${\cal P}(q)$ is the pseudoscalar--meson  with 
momentum $q$, $\eta_\mu$ is the interpolating current for the heavy spin--3/2 baryons
and ${\cal T}$ denotes the time ordering operator. The correlation function in 
Eq. (\ref{estp01}) is calculated in two different ways:
\begin{itemize}
\item in terms of hadronic parameters called the physical or phenomenological
representation,

\item in terms of QCD degrees of freedom by the help of OPE called theoretical or
QCD representation.
\end{itemize}

Matching then these two representations of the same correlation function, we obtain the
QCD sum rules for strong coupling constants. To suppress contributions of the higher
states and continuum, we apply the Borel transformation with respect to the momentum
squared of the initial and final states to both sides of the sum rules. 

We start 
our calculations by considering the physical side. Inserting  complete sets of hadrons 
with the same quantum numbers as the interpolating currents
and isolating the ground states, we obtain
\bea
\label{estp02}
\Pi_{\mu\nu} =  {\lla 0 \vel \eta_{\mu}(0) \ver
B_2(p) \rra \lla B_2(p) {\cal P}(q) \vel \right.  
B_1(p+q) \rra \lla B_1(p+q) \vel \bar{\eta}_{\nu}(0) \ver
0 \rra \over \ga p^2-m_2^2 \dr \left[(p+q)^2-m_1^2\right]}
+ \cdots~,
\eea  
where $\vel B_1(p+q) \rra$ and $\vel B_2(p) \rra$  are the initial and final
spin--3/2 states, and $m_1$ and $m_2$ are their masses, respectively. The dots in
Eq. (\ref{estp02}) represent contributions of the higher states and continuum.
It follows from Eq. (\ref{estp02}) that, in order to
calculate the phenomenological part of the correlation function, the
following matrix elements are needed:
\bea
\label{estp03}
\lla 0 \vel \eta_{\mu}(0) \ver B_2(p) \rra \es \lambda_{B_2} u_{\mu}(p) ~, \nnb \\
\lla B_1(p+q) \vel \bar{\eta}_{\nu}(0) \ver
0 \rra \es\lambda_{B_1} \bar{u}_{\nu}(p+q)~, \nnb \\
\lla B_2(p) {\cal P}(q) \vel \right. B_1(p+q) \rra \es g_{B_1B_2{\cal P}}
\bar{u}_{\alpha}(p)\gamma_5 
u^{\alpha}(p+q)~,
\eea
where $\lambda_{B_1}$ and $\lambda_{B_2}$ are the residues of the initial and final spin--3/2
heavy baryons, $g_{B_1B_2{\cal P}}$ is the strong coupling constant of pseudoscalar
mesons with heavy spin--3/2 baryons and $u_\mu$ is the the Rarita-Schwinger spinor.
Using the above matrix elements and performing summation over the spins of the
Rarita--Schwinger spinors defined as
\bea
\label{ede05}
\sum_s u_\mu (p,s) \bar{u}_\nu (p,s) = - ( {\rlap/p +m } )\Bigg(
- g_{\mu\nu} + {1\over 3} \gamma_\mu \gamma_\nu + {2 p_\mu p_\nu \over 3
m^2} - {p_\mu \gamma_\nu - p_\nu \gamma_\mu \over 3 m} \Bigg)~,
\eea
in principle, one can find the final expression of the correlation function in
phenomenological side. However, the following two principal problems are
unavoidable: 
1) all Lorentz structures are not independent; 2) not only the
heavy spin--$3/2$, but also the heavy spin--$1/2$ states contribute to the physical
side, i.e., the matrix element of the current $\eta_\mu$, sandwiched between the
vacuum and the heavy spin--$1/2$ states, is nonzero and determined in the
following way:
\bea
\label{ede06}
\lla 0 \vel \eta_\mu \ver B(p,s=1/2)\rra = A (4 p_\mu - m
\gamma_\mu) u(p,s=1/2)~,
\eea
where the condition $\gamma_\mu \eta^\mu = 0$ is imposed. To remove the
contribution of the unwanted heavy spin--$1/2$ baryons
and obtain only  independent structures, we  order the Dirac
matrices in a specific way and eliminate the ones that receive
contributions from spin--$1/2$ states. Here, we  choose the
$\gamma_\mu \rlap/p  \rlap/q \gamma_\nu\gamma_5$ ordering of the Dirac
matrices and  obtain the final expression
\bea
\label{ede07}
\Pi_{\mu\nu} \es { \lambda_{B_1} \lambda_{B_2} g_{B_1 B_2 {\cal P}} \over
(p^2-m_2^2) [(p+q)^2 -  m_1^2)]} \Big( g_{\mu\nu}  \rlap/p  \rlap/q \gamma_5+
\mbox{\rm other structures with
$\gamma_\mu$ at the beginning and} \nnb \\
&&\mbox{$\gamma_\nu\gamma_5$ at the end, or terms that are proportional to $(p+q)_{\nu}$
or $p_{\mu}$} \Big)~,
\eea
and to calculate the strong coupling constant $g_{B_1B_2{\cal P}}$, we choose
the structure $g_{\mu\nu}  \rlap/p \rlap/q\gamma_5$, which is free of the unwanted
heavy  spin--$1/2$ states.

Now, we proceed to calculate the correlation function from QCD side. For this aim,
we need to know the explicit expression for the interpolating current of the heavy
spin--3/2 baryons. In constructing the interpolating current for these baryons, we
use the fact that the interpolating current for this case should be symmetric 
with respect to the light quarks. Using this condition, the interpolating current for the  heavy
baryons  with $J=3/2$ containing single heavy quark is written as
\bea
\label{eh32v07}
\eta_\mu = A \epsilon^{abc} \Big\{ (q_1^a C \gamma_\mu q_2^b) Q^c + (q_2^a C
\gamma_\mu Q^b) q_1^c + (Q^a C \gamma_\mu q_1^b) q_2^c \Big\}~,
\eea
where $A$ is the normalization factor, and $a$, $b$ and $c$ are the color indices. In
Table 1, we present the values of $A$ and light quark content of heavy spin-3/2
baryons.
% .........................................................
\begin{table}[thb]

\renewcommand{\arraystretch}{1.3}
\addtolength{\arraycolsep}{-0.5pt}
\small
$$
\begin{array}{|l|c|c|c|c|c|c|}
\hline \hline
 & \Sigma_{b(c)}^{*+(++)} & \Sigma_{b(c)}^{*0(+)} & \Sigma_{b(c)}^{*-(0)}  
 & \Xi_{b(c)}^{*0(+)}    & \Xi_{b(c)}^{*-(0)} 
 & \Omega_{b(c)}^{*-(0)}          \\  \hline
 q_1 & u & u & d & u & d & s \\
 q_2 & u & d & d & s & s & s  \\
 A   & \sqrt{1/3} & \sqrt{2/3} & \sqrt{1/3}
     & \sqrt{2/3} & \sqrt{2/3} & \sqrt{1/3} \\
\hline \hline
\end{array}
$$
\caption{The light quark content $q_1$ and $q_2$ for the heavy baryons with
spin--3/2}
\renewcommand{\arraystretch}{1}
\addtolength{\arraycolsep}{-1.0pt}
\end{table}
% ................................................are.........

Before obtaining the explicit form of the correlation functions on the QCD side, 
we first try to obtain relations among the correlation
functions of the different transitions using some symmetry arguments.
Next, we show that all possible transitions can be described in terms of only one universal
invariant function. We follow the approach given in \cite{azizi1,azizi2,azizi3,azizi4},
where  the coupling constants of heavy spin--1/2 baryons with light  mesons as well as
spin 3/2 baryons--spin 1/2 baryons-- light mesons vertices are calculated (see also 
\cite{Rstp10,Rstp11,Rstp12,Rstp13,Rstp14} for the couplings among light baryons and
light mesons). Here, we should stress that the relations which are presented
below are independent of the choice of Lorentz structures. We start our discussion by 
considering the $\Sigma_b^{*0} \rar \Sigma_b^{*0} \pi^0$ transition. The invariant
function describing  this strong transition can be written in the following general
form:
\bea
\label{estp07}  
\Pi^{\Sigma_b^{*0} \rar \Sigma_b^{*0} \pi^0} = g_{\pi^0\bar{u}u} \Pi_1(u,d,b) +
g_{\pi^0\bar{d}d} \Pi_1^{'}(u,d,b) + g_{\pi^0\bar{b}b} \Pi_2(u,d,b)~,
\eea
where $g_{\pi^0\bar{u}u}$, $g_{\pi^0\bar{d}d}$ and $g_{\pi^0\bar{b}b}$
show coupling of the $\pi^0$ meson to the $\bar{u}u$, $\bar{d}d$ and
$\bar{b}b$ states, respectively.  The interpolating current 
of $\pi^0$ meson is written as
\bea
\label{nolabel} 
J_{\pi^0} = \sum_{u,d} g_{\pi^0 \bar{q}q}\bar{q}\gamma_5 q~, 
\eea
where $g_{\pi^0 \bar{u}u} = -g_{\pi^0 \bar{d}d}={1\over \sqrt{2}}$, $g_{\pi^0
\bar{b}b} = 0$. The invariant functions
$\Pi_1,~\Pi_1^{'}$ and $\Pi_2$ describe the radiation of $\pi^0$
meson from $u,~d$ and $b$ quarks of heavy $\Sigma_b^{*0}$ baryon, respectively, and 
they can formally be defined as:
\bea
\label{estp08}
\Pi_1(u,d,b) \es \lla \bar{u}u \vel \Sigma_b^{*0} \bar{\Sigma}_b^{*0} \ver 0
\rra~, \nnb \\
\Pi_1^{'}(u,d,b) \es \lla \bar{d}d \vel \Sigma_b^{*0} \bar{\Sigma}_b^{*0} \ver 0
\rra~, \nnb \\
\Pi_2(u,d,b) \es \lla \bar{b}b \vel \Sigma_b^{*0} \bar{\Sigma}_b^{*0} \ver 0
\rra~.
\eea
From the interpolating current of $\Sigma^*_b$
baryon, we see that it is symmetric under the exchange $u \lrar d$, so
$\Pi_1^{'}(u,d,b) = \Pi_1(d,u,b)$ and we 
immediately obtain
\bea
\label{estp09}
\Pi^{\Sigma_b^{*0} \rar \Sigma_b^{*0} \pi^0} = {1\over \sqrt{2}} \Big[
\Pi_1(u,d,b) - \Pi_1(d,u,b) \Big]~,
\eea
where under the $SU_2(2)_f$ limit, $\Pi^{\Sigma_b^{*0} \rar
\Sigma_b^{*0} \pi^0} = 0$. Now, we proceed to obtain the invariant function responsible
for other transitions containing the $\pi^0$ meson.  The invariant function for
$\Sigma_b^{*+} \rar \Sigma_b^{*+}
\pi^0$ transition can be obtained from the $\Sigma_b^{*0} \rar \Sigma_b^{*0}
\pi^0$ case by making the replacement $d \rar u$, and using the fact that
$\eta_\mu^{\Sigma_b^{*0}}=\sqrt{2} \eta_\mu^{\Sigma_b^{*+}}$, from which we get,
\bea            
\label{estp10}
4 \Pi_1(u,u,b) = 2 \lla \bar{u}u \vel \Sigma_b^{*+} \bar{\Sigma}_b^{*+} \ver 0
\rra~.
\eea
The factor 4 on the left hand side appears due to the fact that
each $\Sigma^{*+}_b$ contains two $u$ quark, hence there are 4 possible ways for
radiating $\pi^0$ from the $u$ quark.
From Eq. (\ref{estp09}), we get
\bea            
\label{estp11}
\Pi^{\Sigma_b^{*+} \rar \Sigma_b^{*+} \pi^0} = \sqrt{2} \Pi_1(u,u,b).
\eea

Similar arguments lead  to the following result for the 
$\Sigma_b^{*-} \rar \Sigma_b^{*-} \pi^0$ transition:
\bea            
\label{estp12}
\Pi^{\Sigma_b^{*-} \rar \Sigma_b^{*-} \pi^0} =- \sqrt{2} \Pi_1(d,d,b).
\eea
Consider the strong transition $\Xi_b^{*-(0)} \rar \Xi_b^{*-(0)} \pi^0$. The invariant
function for this decay can be obtained from the  $\Sigma_b^{*0} \rar
\Sigma_b^{*0} \pi^0$ case using the fact that $\eta_\mu^{\Xi_b^{*0}} =
\eta_\mu^{\Sigma_b^{*0}}
(d\rar s)$ and  $\eta_\mu^{\Xi_b^{*-}} = \eta_\mu^{\Sigma_b^{*0}}
(u\rar s)$. As a result, we get
\bea
\label{estp13}
\Pi^{\Xi_b^{*0} \rar \Xi_b^{*0} \pi^0} \es {1\over \sqrt{2}}
\Pi_1(u,s,b)~, \nnb \\
\Pi^{\Xi_b^{*-} \rar \Xi_b^{*-} \pi^0} \es - {1\over \sqrt{2}}
\Pi_1(d,s,b)~.
\eea
Now, we proceed to find relations among the invariant functions involving charged
$\pi^\pm$ mesons. We start by considering the
matrix element $\lla \bar{d}d \vel\Sigma_b^{*0} \bar{\Sigma}_b^{*0} \ver 0 \rra$,
where $d$ quarks from the $\Sigma_b^{*0}$ and $\bar{\Sigma}_b^{*0}$ form the
final $\bar{d}d$ state, and $u$ and $b$ quarks are the spectators. The
matrix element $\lla \bar{u}d \vel\Sigma_b^{*+} \bar{\Sigma}_b^{*0} \ver 0
\rra$ explains the case where $d$ quark from $\bar{\Sigma}_b^{*0}$ and $u$ 
quark from $\Sigma_b^{*+}$ form the $\bar{u}d$ state and the remaining $u$ and
$b$ are being again the spectators. From this observations, one expects
that  these matrix elements be proportional to each other. Our calculations 
 support this expectation. Hence,
\bea
\label{estp16}
\Pi^{\Sigma_b^{*0} \rar \Sigma_b^{*+} \pi^-} \es \lla \bar{u}d \vel \Sigma_b^{*+}
\bar{\Sigma}_b^{*0} \ver 0 \rra = \sqrt{2} \lla \bar{d}d \vel \Sigma_b^{*0}
\bar{\Sigma}_b^{*0} \ver 0 \rra = \sqrt{2} \Pi_1(d,u,b)~.
\eea
Making the exchange $u \lrar d$ in Eq. (\ref{estp16}), we obtain
\bea
\label{estp17}  
\Pi^{\Sigma_b^{*0} \rar \Sigma_b^{*-} \pi^+} \es \lla \bar{d}u \vel \Sigma_b^{*-}
\bar{\Sigma}_b^{*0} \ver 0 \rra = \sqrt{2} \lla \bar{u}u \vel \Sigma_b^{*0}
\bar{\Sigma}_b^{*0} \ver 0 \rra = \sqrt{2} \Pi_1(u,d,b)~.        
\eea
Similarly, one can easily show that
\bea
\label{estp170}  
\Pi^{\Sigma_b^{*+} \rar \Sigma_b^{*0} \pi^+} \es \sqrt{2} \Pi_1(d,u,b)~, \nnb \\
 \Pi^{\Sigma_b^{*-} \rar \Sigma_b^{*0} \pi^-} \es \sqrt{2} \Pi_1(u,d,b)~, \nnb \\   
    \Pi^{\Xi_b^{*0} \rar \Xi_b^{*-} \pi^+} \es \Pi_1(d,s,b)~, \nnb \\ 
\Pi^{\Xi_b^{*-} \rar \Xi_b^{*0} \pi^-} \es \Pi_1(u,s,b)~. 
\eea
Calculation of the coupling constants of the members of heavy spin--3/2 baryons to
other pseudoscalar mesons can be done in a similar way as we did for the $\pi^0$ meson.
Here, we shall say that in our calculations, we ignore the mixing between $\eta$ and
$\eta^{'}$ mesons and only consider $\eta_{8}$ instead of physical $\eta$ meson. The interpolating current for $\eta_{8}$ meson has the following form
\bea             
\label{nolabel}
J_{\eta_{8}} = {1\over \sqrt{6}} [ \bar{u}\gamma_5 u + \bar{d}\gamma_5 d
- 2 \bar{s}\gamma_5 s]~,
\eea 
we see that
\bea             
\label{nolabel}
g_{\eta_{8}\bar{u}u} \es g_{\eta_{8}\bar{d}d} = {1\over\sqrt{6}}~,~\mbox{\rm and} 
~~g_{\eta_{8}\bar{s}s} = -{2\over \sqrt{6}}~. 
\eea
For instance, consider
the $\Sigma^{*0}_b \to \Sigma_b^{*0} \eta_{8}$
transition. Following the same lines of calculations as in the $\pi^0$ meson
case, we immediately obtain,
\bea             
\label{eh32v14} 
\Pi^{\Sigma^{*0}_b \to \Sigma_b^{*0} \eta_{8}} = 
{1\over \sqrt{6}} [\Pi_1(u,d,b) + \Pi_1(d,u,b)]~.
\eea             

The invariant function responsible for the $\Xi_b^{*0} \to \Xi_b^{0*} \eta_{8}$ 
transition can be written as:
\bea
\label{eh32v15}
\Pi^{\Xi_b^{*0} \to \Xi_b^{*0} \eta_{8} } \es 
g_{\eta_{8}\bar{u}u} \Pi_1(u,s,b) +
 g_{\eta_{8}\bar{s}s} \Pi_1^{'}(u,s,b) +  g_{\eta_{8}\bar{b}b} \Pi_2(u,s,b) \nnb \\
\es {1\over \sqrt{6}} [\Pi_1(u,s,b) - 2 \Pi_1^{'}(u,s,b)] \nnb \\
\es {1\over \sqrt{6}} [\Pi_1(u,s,b) - 2 \Pi_1(s,u,b)]~.
\eea
For the remaining transitions containing the $\eta_{8}$ meson we obtain
\bea             
\label{eh32v144} 
\Pi^{\Sigma^{*+}_b \to \Sigma_b^{*+} \eta_{8}} \es {2\over \sqrt{6}} \Pi_1(u,u,b) ~,\nnb \\
\Pi^{\Sigma^{*-}_b \to \Sigma_b^{*-} \eta_{8}} \es {2\over \sqrt{6}} \Pi_1(d,d,b) ~,\nnb \\
\Pi^{\Xi_b^{*-} \to \Xi_b^{*-} \eta_{8} }  \es {1\over \sqrt{6}} [\Pi_1(d,s,b) -
2 \Pi_1(s,d,b)]~,\nnb \\
\Pi^{\Omega^{*-}_b \to \Omega_b^{*-} \eta_{8}}\es -{4\over \sqrt{6}} \Pi_1(s,s,b) ~.
\eea  
We also find the following relations for transitions involving $K$ mesons:
\bea             
\label{eh32v1444} 
%1
\Pi^{\Xi_b^{*0} \rar \Sigma_b^{*0} \bar{K}^0 } \es
\Pi^{\Sigma_b^{*0} \rar \Xi_b^{*0}      K^0  } = 
\Pi_1(d,u,b)~, \nnb \\
%3
\Pi^{\Xi_b^{*-} \rar \Sigma_b^{*0}      K^-  } \es
\Pi^{\Sigma_b^{*0} \rar \Xi_b^{*-}      K^+  } = 
\Pi_1(u,d,b)~, \nnb \\
%4
 \Pi^{\Xi_b^{*0} \rar \Sigma_b^{*+}      K^-  } \es
 \Pi^{\Sigma_b^{*+} \rar \Xi_b^{*0}      K^+  }  =
\sqrt{2}\Pi_1(u,u,b)~, \nnb \\
%5
\Pi^{\Omega_b^{*-} \rar \Xi_b^{*0}      K^-  } \es
 \Pi^{\Xi_b^{*0} \rar \Omega_b^{*-}      K^+  } = 
 \Pi^{\Omega_b^{*-} \rar \Xi_b^{*-} \bar{K}^0 } \nnb \\ \es
 \Pi^{\Xi_b^{*-} \rar \Omega_b^{*-}      K^0  } =\sqrt{2} \Pi_1(s,s,b)~, \nnb \\
%6
 \Pi^{\Xi_b^{*-} \rar \Sigma_b^{*-} \bar{K}^0 } \es
 \Pi^{\Sigma_b^{*-} \rar \Xi_b^{*-}      K^0  } =
\sqrt{2} \Pi_1(d,d,b)~, \nnb \\
\eea
The expressions for the charmed baryons can easily be obtained by making the
replacement $b \rar c$ and adding to charge of each baryon a positive unit
charge. 

So far we have obtained all possible strong transitions among the
heavy spin--3/2 baryons  with pseudoscalar mesons that
are described in terms of only one invariant function
$\Pi_1$. This function can be calculated in deep Euclidean region, where $-p^2 \rar +\infty$
and $-(p+q)^2 \rar +\infty$ using the OPE in terms of DA's of
the pseudoscalar mesons as well as light and heavy quark propagators. In obtaining the
expression of $\Pi_1$ in QCD side, the nonlocal matrix elements of types $\lla P(q)
\vel \bar{q}(x) \Gamma q(0) \ver 0
\rra$ and $\lla P(q) \vel \bar{q}(x) G_{\mu\nu} q(0) \ver 0 \rra$ appear, where
$\Gamma$ is any arbitrary Dirac matrix. Up to twist--4
accuracy, these matrix elements are determined in terms of the DA's 
of the pseudoscalar mesons. These matrix elements as well as the explicit
expressions of DA's are given in \cite{Rstp15,Rstp16,Rstp17}.

In  calculation of the invariant function $\Pi_1$, we
also need to know the expressions of the light and heavy quark propagators.
The light quark propagator, in the  presence of an external gluon field, is calculated
in \cite{Rh32v21}:
\bea
\label{eh32v18}
S_q(x) \es {i \rlap/x\over 2\pi^2 x^4} - {m_q\over 4 \pi^2 x^2} -
{\lla \bar q q \rra\over 12} \left(1 - i {m_q\over 4} \rlap/x \right) -
{x^2\over 192} m_0^2 \lla \bar q q \rra  \left( 1 -
i {m_q\over 6}\rlap/x \right) \nnb \\
&&  - i g_s \int_0^1 du \left[{\rlap/x\over 16 \pi^2 x^2} G_{\mu \nu} (ux)
\sigma_{\mu \nu} - {i\over 4 \pi^2 x^2} u x^\mu G_{\mu \nu} (ux) \gamma^\nu
\right. \nnb \\
&& \left.
 - i {m_q\over 32 \pi^2} G_{\mu \nu} \sigma^{\mu
 \nu} \left( \ln \left( {-x^2 \Lambda^2\over 4} \right) +
 2 \gamma_E \right) \right]~,
\eea
where $\gamma_E \simeq 0.577$ is the Euler constant, and $\Lambda$ is the
scale parameter. In further numerical calculations, we choose it as
$\Lambda=(0.5 \div 1)~GeV$ (see \cite{Rh32v22,Rh32v23}).
The heavy quark propagator in an external gluon field is given as:
\bea
\label{eh32v19}
S_Q(x) = S_Q^{free}(x) -
ig_s \int {d^4k \over (2\pi)^4} e^{-ikx} \int_0^1
du \Bigg[ {\rlap/k+m_Q \over 2 (m_Q^2-k^2)^2} G^{\mu\nu} (ux)
\sigma_{\mu\nu} +
{u \over m_Q^2-k^2} x_\mu G^{\mu\nu} \gamma_\nu \Bigg]~,
\eea
where $S_Q^{free}(x)$ is the free heavy quark operator in coordinate space and it 
 is given by
\bea
\label{nolabel}
S_Q^{free} (x) = 
{m_Q^2 \over 4 \pi^2} {K_1(m_Q\sqrt{-x^2}) \over \sqrt{-x^2}} -
i {m_Q^2 \rlap/{x} \over 4 \pi^2 x^2} K_2(m_Q\sqrt{-x^2})~, 
\eea
where $K_1$ and $K_2$ are the modified Bessel function of the second kind.

Using the explicit expressions of the heavy and light quark  propagators and
definitions of DA's for the pseudoscalar mesons, we calculate the correlation
function from the QCD side.
Equating the coefficients of the structure $g_{\mu\nu}\rlap/p \rlap/q \gamma_5$
from both sides of the correlation function  and applying the Borel transformation
with respect to the variables $p^2$ and $(p+q)^2$ in order to suppress the
contributions of the higher states and continuum, we get the following sum rules
for the strong coupling constants of the pseudoscalar mesons with heavy spin--3/2
baryons:
\bea
\label{estp20}
g_{B_1 B_2 {\cal P}}  = {1 \over \lambda_{B_1} \lambda_{B_2}} e^{{m_1^{2} 
\over M_1^2} + {m_2^{2}
\over M_2^2} }\, \Pi_1~,
\eea
where $M_1^2$ and $M_2^2$ are the Borel mass parameters correspond to
the initial and  final heavy baryons, respectively. The contributions of the 
higher states and continuum  are obtained by invoking the duality
condition, which means that above the thresholds $s_1$ and $s_2$ the double
spectral density $\rho^h (s_1,s_2)$ coincides with the spectral density
derived from QCD side of the correlation function. The procedure for
obtaining double spectral density from QCD side and subtraction of higher
states and continuum contributions are explained in detail in \cite{R23},
which we use in the present work. 

The masses of the initial and final
baryons are  equal to each other, so we take $M_1^2=M_2^2=2 M^2$
and the residues $\lambda_{B_1}$ and $\lambda_{B_2}$   are calculated in \cite{Rstp21}.
The explicit expression for $\Pi_1$ function is quite lengthy and we do not
present its explicit form here.

\section{Numerical results}

This section is devoted to the numerical analysis of the sum rules for strong
coupling constants of pseudoscalar mesons with  heavy spin--3/2
baryons. The main input
parameters of these LCSRs are DA's of the pseudoscalar mesons which are given in
\cite{Rstp15,Rstp16,Rstp17}. Some other input parameters entering to
the sum rules are $\lla \bar{q}q\,(2~GeV)\rra = -(274_{-17}^{+15}~MeV)^3$ and
$\mu_\pi (2\,GeV)=f_\pi m_\pi^2/(m_u+m_d) =( 2.43 \pm 0.42)\,GeV$ \cite{R25},
$\langle \bar s s\rangle=0.8\uu$,
$\langle0\mid {1\over \pi} \alpha_sG^2\mid 0\rangle=(0.012\pm0.004)~GeV^4$,
$m_0^2 = (0.8 \pm 0.2)~GeV^2$ \cite{Rstp22}, $f_\pi = 0.131~GeV$, $f_K = 0.16~GeV$ and 
$f_{\eta_{8}} = 0.13~GeV$ \cite{Rstp15}. 

\begin{figure}[h]
\begin{center}
\scalebox{0.8}{\includegraphics{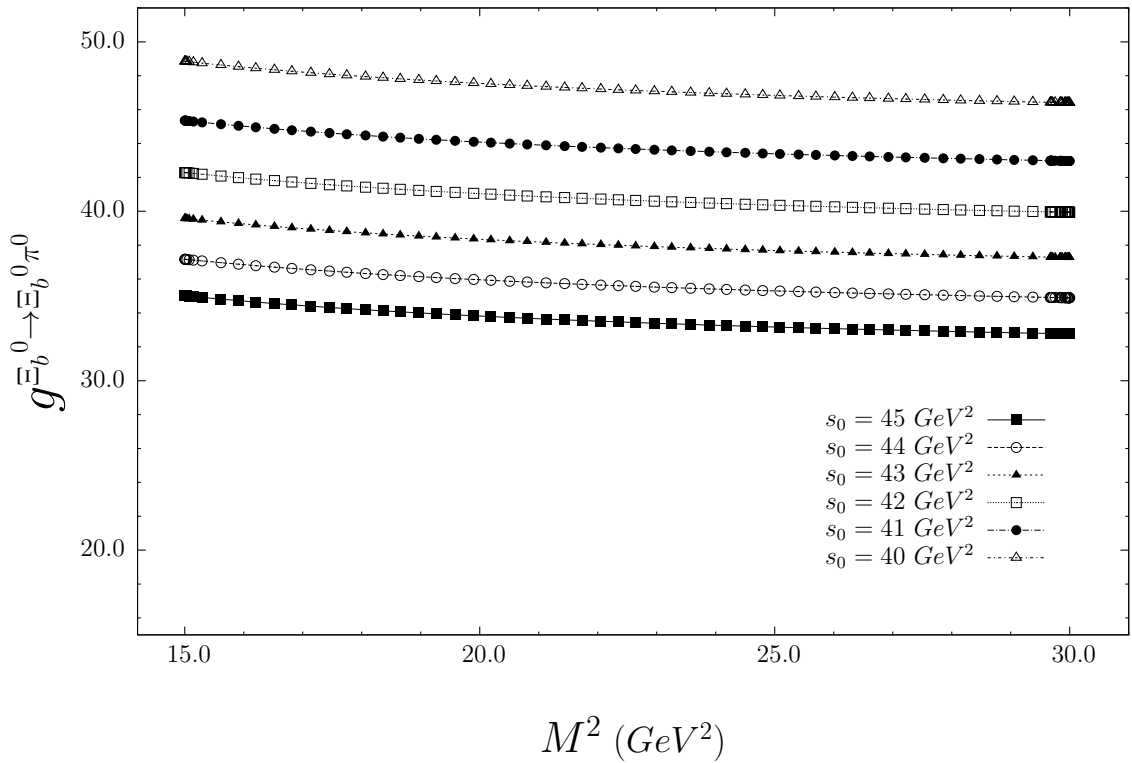}}
\end{center}
\caption{The dependence of the strong coupling constant for 
the $\Xi_b^{*0} \rar \Xi_b^{*0} \pi^0$ transition at several 
 fixed values of  $s_0$.}
\end{figure}

The sum rules for the strong coupling constants include also two
auxiliary parameters: Borel mass parameter $M^2$ and  continuum threshold $s_0$.
These are not physical quantities, hence the result for  coupling constants
should be independent of them. Therefore, we should look for working
regions of these parameters, where coupling constants remain approximately
unchanged. The upper limit of $M^2$ is obtained requiring that the contribution of
the higher states and continuum is small and constitutes only few percent of the
total dispersion integral. The lower bound of   $M^2$ is obtained demanding that the
series of the light cone expansion with increasing twist should be convergent.
These  conditions lead to the  working region $15~GeV^2 \le M^2 \le 30~GeV^2$ for the
bottom heavy spin--3/2 baryons and $4~GeV^2 \le M^2 \le 10~GeV^2$ for the charmed
cases. In these intervals, the twist--4 contributions does not exceed (4--6)\%
of the total result. Our analysis also shows that contribution of the higher states and continuum 
is less than 25\%. The continuum threshold $s_0$ is not totally arbitrary but it is correlated
to the energy of the first excited state with the same quantum numbers as the interpolating
current. Our calculations show that in the interval $(m_B + 0.4)^2~GeV^2\leq 
s_0\leq(m_B + 0.8)^2~GeV^2$, the strong coupling constants weakly depend on this parameter. 

As an example, let us consider the
$\Xi_b^{*0} \rar \Xi_b^{*0} \pi^0$ transition. The dependence of
the strong coupling constant for the $\Xi_b^{*0} \rar \Xi_b^{*0} \pi^0$
transition on $M^2$ at different fixed values of the $s_0$ is depicted in Fig. (1). 
From this figure, we see that the strong coupling constant for $\Xi_b^{*0} \rar \Xi_b^{*0}
\pi^0$ shows a good stability in the ``working region" of $M^2$.  This figure also depicts
that the result of strong coupling constant has weak dependency on the continuum threshold
in its working region. From this figure, we deduce  $g_{\Xi_b^{*0} \Xi_b^{*0} \pi^0}=41\pm7$.
From the same manner, we analyze all considered strong vertices and obtain the numerical
values as presented in Table (2). Note that, in this Table, we show only those couplings
which could not be obtained by the SU(2) symmetry rotations.
\begin{table}[t]

\renewcommand{\arraystretch}{1.3}
\addtolength{\arraycolsep}{-0.5pt}
\small
$$
\begin{array}{|l|r@{\pm}lr@{}l||l|r@{\pm}lr@{}l|}
\hline \hline  
 \multirow{2}{*}{}        &\multicolumn{4}{c||}{\mbox{Bottom Baryons}}   &  
 \multirow{2}{*}{}        &\multicolumn{4}{c|}{\mbox{Charmed Baryons}}    \\ \hline
 g^{\Xi_b^{*0}     \rar \Xi_b^{*0}     \pi^0}     &~~~~~~~41&9    ~(4.0\pm 0.5)    && &
 g^{\Xi_c^{*+}     \rar \Xi_c^{*+}     \pi^0}     &~~~~~~ 22&6    ~(5.3\pm 0.6)    &&   \\ 
 g^{\Sigma_b^{*0}  \rar \Sigma_b^{*-}  \pi^+}     &       75&15   ~(8.0\pm 0.7)    && &
 g^{\Sigma_c^{*+}  \rar \Sigma_c^{*0}  \pi^+}     &       42&10    ~(11.0\pm 0.7)   &&   \\
 g^{\Xi_b^{*0}     \rar \Sigma_b^{*+}   K^-}      &       65&14   ~(7.8\pm 0.8)    && &
 g^{\Xi_c^{*+}     \rar \Sigma_c^{*++}  K^-}      &       37&8    ~(10.2\pm 0.6)   &&   \\
 g^{\Omega_b^{*-}  \rar \Xi_b^{*0}      K^-}      &       73&16   ~(8.0\pm 0.8)    && &
 g^{\Omega_c^{*0}  \rar \Xi_c^{*+}      K^-}      &       37&9    ~(9.8\pm 0.8)    &&   \\
 g^{\Sigma_b^{*+}  \rar \Sigma_b^{*+}  \eta_{8}}    &       45&10    ~(5.8\pm 0.6)    && &
 g^{\Sigma_c^{*++} \rar \Sigma_c^{*++} \eta_{8}}    &       25&6    ~(7.8\pm 0.6)    &&   \\
 g^{\Xi_b^{*0}     \rar \Xi_b^{*0}     \eta_{8}}    &      19&4    ~(2.7\pm 0.3)    && &
 g^{\Xi_c^{*+}     \rar \Xi_c^{*+}     \eta_{8}}    &    11.5&1.8  ~(3.2\pm 0.4)    &&   \\
 g^{\Omega_b^{*-}  \rar \Omega_b^{*-}  \eta_{8}}    &      95&22   ~(11.5\pm 1.8)   && &
 g^{\Omega_c^{*0}  \rar \Omega_c^{*0}  \eta_{8}}    &      52&10    ~(13.6\pm 0.7)   &&   \\
 \hline \hline
\end{array}
$$
\caption{The values of the strong coupling constants  for the transitions
among the  heavy spin--3/2 baryons with pseudoscalar mesons.}

\renewcommand{\arraystretch}{1}
\addtolength{\arraycolsep}{-1.0pt}

\end{table}
The errors in the values of the coupling constants presented in the
Table  include uncertainties coming from the variations of the
$s_0$ and $M^2$ as well as those coming from the other input parameters. Here, we should stress that in the
Table (2), we only present the modules of the strong coupling constants, since the sum
rules approach can not predict the sign of the residues of the heavy baryons. Our numerical calculations show that  the HQET  are violated approximately  5\% (16\%) for the 
coupling constants of the heavy baryons containing $b$($c$) quark.  Finally, we also check the $SU(3)_f$
symmetry violating effects and see that they change the results maximally
about 8\%.

At the end of this section, it should be mentioned  that the predictions of the sum rules on
light baryon-meson couplings  strongly depend on the choice of the
structure (for more detail see \cite{R27}). In connection with this
point, here immediately arises the question whether or not a similar
situation occurs for the case of the heavy baryon-light meson couplings. In
order to answer this question, we also analyze the coupling constants predicted by
the $g_{\mu\nu} \gamma_5$ structure, which are presented in Table (2)
(see the values in the  brackets). From these results, it follows  that the values
of the strong coupling constants of the heavy hadrons containing $b$($c$)
quark with light pseudoscalar mesons decrease by a factor of about 8(4) times for
the $g_{\mu\nu} \gamma_5$ structure. However, sticking on the same criteria as is used in
\cite{R27}, we find that the $g_{\mu\nu} \rlap/{p}\rlap/{q} \gamma_5$ is a more pertinent 
Dirac structure.

In conclusion, the strong coupling constants of light pseudoscalar mesons with
heavy spin--3/2 baryons have been studied within LCSR.  Using symmetry
arguments, the Lorenz structure independent  relations among different correlation
functions have been obtained. It has been  shown that all possible transitions
can be written in terms of one universal invariant function. Furthermore, it
has been observed that the values of the coupling constants are strongly structure
dependent similar to the case of light baryon-meson couplings.
The numerical values
of those strong coupling constants which could not be obtained via the
SU(2) symmetry rotations have been also presented.
\newpage

\bAPP{}{}

In this appendix, we present some details of our calculations, i.e. how we perform the Fourier and Borel transformations as well as continuum subtraction. For this aim, 
let us consider the following generic term:                                                                                                                                                                     
\baeeq
\label{nolabel}
T=\int d^4x ~e^{ipx} \int_0^1 du~ e^{iuqx} f(u) \frac{K_\nu(m_Q\sqrt{-x^2})}{(\sqrt{-x^2})^n},
\eaeeq
where $K_\nu$ is the modified Bessel function of order $\nu$ appearing in the propagator of heavy quark. Using the integral representation of the modified Bessel function 
\baeeq
 K_\nu(m_Q\sqrt{-x^2})=\frac{\Gamma(\nu+1/2)2^\nu}{\sqrt{\pi}m_Q^\nu}\int_0^\infty dt~cos(m_Qt)\frac{(\sqrt{-x^2})^\nu}{(t^2-x^2)^{\nu+1/2}},
\eaeeq
we get
\baeeq
T=\int d^4x\int_0^1 du~e^{iPx}f(u)\frac{\Gamma(\nu+1/2)2^\nu}{\sqrt{\pi}m_Q^\nu}\int_0^\infty dt~cos(m_Qt)\frac{1}{(\sqrt{-x^2})^{n-\nu}(t^2-x^2)^{\nu+1/2}},
\eaeeq
where $P=p+uq$. For further calculations we go to the Euclidean space. Using the identity
\baeeq
\frac{1}{Z^n}=\frac{1}{\Gamma(n)}\int_0^\infty d\alpha~\alpha^{n-1} e^{-\alpha Z},
\eaeeq
we have
\baeeq
T=\frac{-i2^\nu}{\sqrt{\pi}m_Q^\nu\Gamma(\frac{n-\nu}{2})}\int_0^1 du f(u)\int_0^\infty dt~e^{im_Qt}\int_0^\infty dy ~y^{\frac{n-\nu}{2}-1}
\int_0^\infty dv ~v^{\nu-\frac{1}{2}}e^{-vt^2}\int d^4\tilde{x}e^{-i\tilde{P}\tilde{x}-y\tilde{x}^2-v\tilde{x}^2},\nnb\\
\eaeeq
where $\tilde{}$ means vectors in Euclidean space and we will consider only the real part of the complex exponential function, $e^{im_Qt}$. After performing Gaussian integral
 over $\tilde{x}$, we obtain
\baeeq
T=\frac{-i2^\nu\pi^2}{\sqrt{\pi}m_Q^\nu\Gamma(\frac{n-\nu}{2})}\int_0^1 du f(u)\int_0^\infty dt~e^{im_Qt}\int_0^\infty dy ~y^{\frac{n-\nu}{2}-1}
\int_0^\infty dv ~v^{\nu-\frac{1}{2}}e^{-vt^2}\frac{e^{-\frac{\tilde{P}^2}{4(y+v)}}}{(y+v)^2}.
\eaeeq
The next step is to perform the integration over $t$. As a result we obtain
\baeeq
T=\frac{-i2^\nu\pi^2}{m_Q^\nu\Gamma(\frac{n-\nu}{2})}\int_0^1 du f(u) \int_0^\infty dy ~y^{\frac{n-\nu}{2}-1}
\int_0^\infty dv ~v^{\nu-1}e^{-\frac{m_Q^2}{4v}}\frac{e^{-\frac{\tilde{P}^2}{4(y+v)}}}{(y+v)^2}.
\eaeeq
Let us define new variables,
\baeeq
~~~~~~~~~~~~~~~~~~~~~~~~~\lambda=v+y,~~~~~~~~~~~~~\tau=\frac{y}{v+y},
\eaeeq
hence
\baeeq
T=\frac{-i2^\nu\pi^2}{m_Q^\nu\Gamma(\frac{n-\nu}{2})}\int_0^1 du f(u) \int d\lambda \int d\tau
 ~ \lambda^{\frac{n+\nu}{2}-3} \tau^{\frac{n-\nu}{2}-1}(1-\tau)^{\nu-1}e^{-\frac{m_Q^2}{4\lambda(1-\tau)}} e^{-\frac{\tilde{P}^2}{4\lambda}}
\eaeeq
Performing Double Borel transformation with respect to the $\tilde{p}^2$ and  $(\tilde{p}+\tilde{p})^2$ using the 
\baeeq
{\cal B}(M^2)e^{-\alpha p^2}=\delta(1/M^2-\alpha),
\eaeeq
we get
\baeeq
{\cal B}(M_1^2){\cal B}(M_2^2)T&=&\frac{-i2^\nu\pi^2}{m_Q^\nu\Gamma(\frac{n-\nu}{2})}\int_0^1 du f(u) \int d\lambda \int d\tau
 ~\lambda^{\frac{n+\nu}{2}-3} \tau^{\frac{n-\nu}{2}-1}(1-\tau)^{\nu-1}e^{-\frac{m_Q^2}{4\lambda(1-\tau)}}e^{-\frac{u(u-1)\tilde{q}^2}{4\lambda}}\nnb\\
&\times&
\delta(\frac{1}{M_1^2}-\frac{u}{4 \lambda})\delta(\frac{1}{M_2^2}-\frac{1-u}{4 \lambda}).
\eaeeq
Performing integrals over $u$ and $\lambda$, we obtain
\baeeq
{\cal B}(M_1^2){\cal B}(M_2^2)T&=&\frac{-i2^\nu4^2\pi^2}{m_Q^\nu\Gamma(\frac{n-\nu}{2})} \int d\tau f(u_0) 
 \Bigg(\frac{M^2}{4}\Bigg)^{\frac{n+\nu}{2}}\tau^{\frac{n-\nu}{2}-1}(1-\tau)^{\nu-1}
e^{-\frac{m_Q^2}{M^2(1-\tau)}}e^{\frac{\tilde{q}^2}{M_1^2+M_2^2}},\nnb\\
\eaeeq
where, $u_0=\frac{M_2^2}{M_1^2+M_2^2}$ and $M^2=\frac{M_1^2M_2^2}{M_1^2+M_2^2}$. Replacing $\tau=x^2$, we will have
\baeeq
{\cal B}(M_1^2){\cal B}(M_2^2)T&=&\frac{-i2^{\nu+1}4^2\pi^2}{m_Q^\nu\Gamma(\frac{n-\nu}{2})} \int_0^1 dx f(u_0) 
 \Bigg(\frac{M^2}{4}\Bigg)^{\frac{n+\nu}{2}}x^{n-\nu-1}(1-x^2)^{\nu-1}
e^{-\frac{m_Q^2}{M^2(1-x^2)}}e^{\frac{\tilde{q}^2}{M_1^2+M_2^2}}.\nnb\\
\eaeeq
Finally, after changing the variable $\eta=\frac{1}{1-x^2}$ and using $q^2=m^2_{{\cal P}}$,  we get
\baeeq
{\cal B}(M_1^2){\cal B}(M_2^2)T&=&\frac{-i2^{\nu+1}4^2\pi^2}{m_Q^\nu\Gamma(\frac{n-\nu}{2})}  f(u_0) 
 \Bigg(\frac{M^2}{4}\Bigg)^{\frac{n+\nu}{2}}e^{-\frac{m^2_{{\cal P}}}{M_1^2+M_2^2}}\Psi\Bigg(\alpha,\beta,\frac{m_Q^2}{M^2}\Bigg),
\eaeeq
where
\baeeq
\Psi\Bigg(\alpha,\beta,\frac{m_Q^2}{M^2}\Bigg)=\frac{1}{\Gamma(\alpha)}\int_1^\infty d\eta e^{-\eta\frac{m_Q^2}{M^2}} \eta^{\beta-\alpha-1}(\eta-1)^{\alpha-1},
\eaeeq
with $\alpha=\frac{n-\nu}{2}$ and $\beta=1-\nu$.

Now, let us discuss how  contribution of the continuum and  higher states are subtracted. For this aim we consider a generic term of the
 form 
\baeeq
A=(M^2)^n f(u_0) \Psi\Bigg(\alpha,\beta,\frac{m_Q^2}{M^2}\Bigg).
\eaeeq
 We should find the spectral density corresponding to this term (see also \cite{Beilin}). The first step is to expand $f(u_0)$ as
\baeeq
f(u_0)=\Sigma a_ku_0^k.
\eaeeq
As a result we get
\baeeq
A=\Bigg(\frac{M_1^2 M_2^2}{M_1^2+M_2^2}\Bigg)^n\Sigma a_k\Bigg(\frac{ M_2^2}{M_1^2+M_2^2}\Bigg)^k\frac{1}{\Gamma(\alpha)}\int_1^\infty d\eta 
e^{-\eta\frac{m_Q^2}{M^2}} \eta^{\beta-\alpha-1}(\eta-1)^{\alpha-1}.
\eaeeq
Introducing new variables, $\sigma_1=\frac{1}{M_1^2}$ and $\sigma_2=\frac{1}{M_2^2}$, we have
\baeeq
A&=&\Sigma a_k \frac{\sigma_1^k}{(\sigma_1+\sigma_2)^{n+k}}\frac{1}{\Gamma(\alpha)}\int_1^\infty d\eta 
e^{-\eta m_Q^2(\sigma_1+\sigma_2)} \eta^{\beta-\alpha-1}(\eta-1)^{\alpha-1}\nnb\\
&=&\Sigma a_k \frac{\sigma_1^k}{\Gamma(n+k)\Gamma(\alpha)}\int_1^\infty d\eta 
 e^{-\eta m_Q^2(\sigma_1+\sigma_2)}\eta^{\beta-\alpha-1}(\eta-1)^{\alpha-1}\int_0^\infty d\xi e^{-\xi(\sigma_1+\sigma_2)}\xi^{n+k-1}\nnb\\
&=&\Sigma a_k \frac{\sigma_1^k}{\Gamma(n+k)\Gamma(\alpha)}\int_1^\infty d\eta 
 \eta^{\beta-\alpha-1}(\eta-1)^{\alpha-1}\int_0^\infty d\xi \xi^{n+k-1}e^{-(\xi+\eta m_Q^2)(\sigma_1+\sigma_2)}\nnb\\
&=&\Sigma a_k \frac{(-1)^k}{\Gamma(n+k)\Gamma(\alpha)}\int_1^\infty d\eta 
 \eta^{\beta-\alpha-1}(\eta-1)^{\alpha-1}\int_0^\infty d\xi \xi^{n+k-1}\Bigg((\frac{d}{d\xi})^ke^{-(\xi+\eta m_Q^2)\sigma_1}\Bigg)e^{-(\xi+\eta m_Q^2)\sigma_2}.\nnb\\
\eaeeq
Applying double Borel transformation with respect to $\sigma_1\rar\frac{1}{s_1}$ and $\sigma_2\rar\frac{1}{s_2}$, we obtain the  spectral density
\baeeq
\rho(s_1,s_2)&=&\Sigma a_k \frac{(-1)^k}{\Gamma(n+k)\Gamma(\alpha)}\int_1^\infty d\eta 
 \eta^{\beta-\alpha-1}(\eta-1)^{\alpha-1}\int_0^\infty d\xi \xi^{n+k-1}\Bigg((\frac{d}{d\xi})^k\delta(s_1-(\xi+\eta m_Q^2))\Bigg)\nnb\\
&\times&\delta(s_2-(\xi+\eta m_Q^2)).
\eaeeq
Performing integration over $\xi$, finally we obtain the following expression for the double spectral density:
\baeeq
\rho(s_1,s_2)&=&\Sigma a_k \frac{(-1)^k}{\Gamma(n+k)\Gamma(\alpha)}\int_1^\infty d\eta 
 \eta^{\beta-\alpha-1}(\eta-1)^{\alpha-1}(s_1-\eta m_Q^2)^{n+k-1}\Bigg((\frac{d}{ds_1})^k\delta(s_2-s_1)\Bigg)\nnb\\
&\times&\theta(s_1-\eta m_Q^2),
\eaeeq
or
\baeeq
\rho(s_1,s_2)&=&\Sigma a_k \frac{(-1)^k}{\Gamma(n+k)\Gamma(\alpha)}\int_1^{s_1/m_Q^2} d\eta 
 \eta^{\beta-\alpha-1}(\eta-1)^{\alpha-1}(s_1-\eta m_Q^2)^{n+k-1}\Bigg((\frac{d}{ds_1})^k\delta(s_2-s_1)\Bigg).\nnb\\
\eaeeq
Using this spectral density, the continuum subtracted correlation function in the Borel scheme corresponding to the considered term can be written as:
\baeeq
\Pi^{sub}=\int_{m_Q^2}^{s_0}ds_1\int_{m_Q^2}^{s_0}ds_2~\rho(s_1,s_2)e^{-s_1/M_1^2}e^{-s_2/M_2^2}.
\eaeeq
Defining new variables, $s_1=2 s v$ and $s_2=2 s(1- v)$, we get
\eAPP
\baeeq
\Pi^{sub}=\int_{m_Q^2}^{s_0}ds\int dv~\rho(s_1,s_2)(4s)e^{-2sv/M_1^2}e^{-2s(1-v)/M_2^2}.
\eaeeq
Using the expression for the spectral density, one can get
\baeeq
\Pi^{sub}&=&\Sigma a_k \frac{(-1)^k}{\Gamma(n+k)\Gamma(\alpha)}\int_{m_Q^2}^{s_0}ds\int dv\frac{1}{2^ks^k}\Bigg((\frac{d}{dv})^k\delta(v-1/2)\Bigg)\nnb\\
&\times&\int_1^{2sv/m_Q^2} d\eta ~
 \eta^{\beta-\alpha-1}(\eta-1)^{\alpha-1}(2sv-\eta m_Q^2)^{n+k-1}e^{-2sv/M_1^2}e^{-2s(1-v)/M_2^2}.
\eaeeq
Integrating over $v$, finally we obtain
\baeeq
\Pi^{sub}&=&\Sigma a_k \frac{(-1)^k(-1)^k}{\Gamma(n+k)\Gamma(\alpha)}\int_{m_Q^2}^{s_0}ds\frac{1}{2^ks^k}\nnb\\
&\times&\Bigg[(\frac{d}{dv})^k\int_1^{2sv/m_Q^2} d\eta ~
 \eta^{\beta-\alpha-1}(\eta-1)^{\alpha-1}(2sv-\eta m_Q^2)^{n+k-1}e^{-2sv/M_1^2}e^{-2s(1-v)/M_2^2}\Bigg]_{v=1/2}.\nnb\\
\eaeeq

\end{document}